\documentclass[aps,pra,reprint,superscriptaddress,10pt,showpacs,a4paper]{revtex4-1}
\usepackage{amsmath,amssymb,amsfonts}
\usepackage{mathtools}
\usepackage{graphicx}
\usepackage{txfonts}
\usepackage{color, xcolor}

\usepackage[unicode,breaklinks]{hyperref}
\hypersetup{
    unicode=true,
    a4paper=true,
    plainpages=false,
    pdftitle={Learning the Position of Image Vortices from Data},
    pdfauthor={Ryan Doran},
    pdfsubject={Learning the Position of Image Vortices from Data},
    colorlinks=true,
    linkcolor=blue,
    citecolor=blue,
    filecolor=black,
    urlcolor=blue
}
\usepackage{tikz}

\newcommand{\rr}{\mathbf{r}}
\newcommand{\vv}{\mathbf{v}}

\begin{document}
\title{Learning the Position of Image Vortices from Data}

\author{Ryan Doran\address{ryan.doran@newcastle.ac.uk}}
\email[]{ryan.doran@newcastle.ac.uk}
\affiliation{Joint Quantum Centre (JQC) Durham--Newcastle, School of Mathematics, Statistics and Physics, Newcastle University, Newcastle upon Tyne, NE1 7RU, UK}
\date{\today}

\begin{abstract}
  The point vortex model is an idealized model for describing the dynamics of many vortices with numerical efficiency, and has been shown to be powerful in modeling the dynamics of vortices in a superfluid. The model can be extended to describe vortices in fluids with a well defined boundary, as an image vortex can be added to the equations of motion to impose the correct velocity profile at the boundary. The mathematical formulation of the image vortex depends on the boundary in question, and is well known for a wide variety of problems, although the formulation of an image vortex in a fluid with a soft boundary remains under debate, as the boundary condition is ill-posed. Such a boundary is common-place in the dynamics of a vortex in an ultra-cold atomic Bose-Einstein condensate, for example, which is typically trapped in a harmonic potential. 
  
  In order to address this problem, the Sparse Identification of Nonlinear Dynamics framework is applied to data from mean-field simulations to extract an approximate point vortex model for a vortex in a circular power law trapping potential. A formulation for the position of an image vortex in such a trap is presented, and the accuracy of this model is evaluated.
\end{abstract}

\maketitle

\section{Introduction}

Vortices- localised areas of fluid that flow around an axis- are the fundamental building blocks of turbulence, a phenomenon that we experience every day. They have been observed on a range of length scales, from vortex cores widths of a few billionths of a meter in superfluid $^4\mathrm{He}$ \cite{Barenghi2014} to vortex rings in a supernova that are $10^{12}$ meters across \cite{Wadas2024-astro}, and arise in systems as diverse as non-linear optics \cite{Arecchi1991}, polariton fluids \cite{Dominici2018}, to volcanic vortex rings \cite{Chacraborty2009_volcano} and even the collective dynamics of crowds \cite{Gu2025_human}. There is a rich history of the study of vortex dynamics that spans over 150 years \cite{Helmholtz1858,Helmholtz1867,Thomson1869}.
 
The Point Vortex Model is an idealised model that described the motion of vortices in the velocity field of other vortices in the system, in the limit of infintesimally small cores. The Hamiltonian formulation of the problem was presented by Kirchhoff \cite{Kirchhoff1883}, for vortices in an unbounded plane. This was later developed to consider vortices within bounded domains \cite{Lin1941motion,Lin1943} and vortices on curved surfaces \cite{hally1980stability}. Despite being a relatively simple model, the point vortex model has led to a variety of remarkable results:   Onsager realised that bounded systems of point vortices could lead to states with decreasing entropy \cite{Onsager1949}, an idea that was later developed in the context of 2D turbulence \cite{Kraichnan1980}, leading to the formation of Onsager Vortex Clusters \cite{Johnstone2019} and negative-temperature states \cite{Gauthier2019}. In classical fluids, the point vortex model is limited by the varying size of vortex cores and arbitrary circulation of the fluid. Superfluids of ultra-cold atomic Bose-Einstein Condensates, however, are the perfect realisation of the point vortex model; their embodiment as an ideal fluid gives rise to quantization of circulation and the size of the vortex core is fixed by the atomic species \cite{Matthews1999,Anderson2000}, while vortices reconnect in well-defined events \cite{Serafini2017}. The faithfulness with which the point vortex model reproduces superfluid vortex dynamics has produced a number of results in the field, including predictions about the ordering \cite{Simula2014} and scaling laws \cite{Billam2015,Reeves2017} of turbulent superfluids, and the prediction of dynamics in experiments comprising few \cite{Yauhen2019} and many (turbulent) \cite{Reeves2022} vortices. While the dynamics of superfluid vortices are well predicted by the point vortex model in homogeneous systems (such as theoretical systems with periodic boundary conditions \cite{Weiss1991}) or systems with a hard wall trap that give rise to a well-defined fluid boundary \cite{Gauthier2016}, where one can ensure that there is no fluid flow across the boundary by using the method of images \cite{Newton}. It is problematic, however, to use the point vortex to describe vortices in a harmonic trapping potential, which is the most typical trap for a superfluid formed of a BEC. Here, the ``edge'' of the fluid is poorly defined, as the density smoothly decreases from its peak value at the centre of the trap and vanishes at the near the edge. As a consequence, it is unclear where an image vortex should be placed.

Data continues to play a key role in the model discovery of physical systems, since it was shown that models can be derived automatically from observations  \cite{Bongard2007,Schmidt2009}. Recently, sparsity-promoting techniques such as the Sparse Identification of Non-linear Dynamics (SINDy) framework have have been developed to identify models from data - balancing model accuracy against overfitting. SINDy has been applied to a number of problems in fluid mechanics, spanning systems governed by ordinary differential equations: the time history of principal orthogonal decomposition (POD) coefficients in vortex shedding past a cylinder \cite{Brunton2016}, coefficients of 2 active DMD modes in a shear-driven cavity flow \cite{Callaham2022}, and the dynamics of dominant POD coefficients in 3D magnetohydrodynamics \cite{Kaptanoglu2021}; as well as systems governed by partial differential equations such as 3D turbulent channel flow \cite{Gurevich2024}. Since its introduction \cite{Brunton2016}, several variations of SINDy have been developed including SINDy with control \cite{Kaiser2018,Fasel2021} (for observations in the low-data high-noise limit), SINDy for rational functions \cite{Mangan2016},  SINDy for integral equations \cite{Schaeffer2017}, weak form SINDy \cite{Reinbold2020,Messenger2021}, SINDy on ensembles of data \cite{Fasel2022,Hirsh2022,Gao2023,Fung2025} utilizing a Bayesian framework \cite{Hirsh2022,Gao2023,Fung2025}, and SINDy using automatic regression \cite{Egan2024}. 

In this paper the SINDy framework is applied to the problem of a single vortex in a superfluid gas confined by a power-law trapping potential. To begin, the known case of a single vortex in a hard wall trap is examined, where data of the vortex trajectory can be obtained by evolving the point vortex model, and the SINDy algorithm correctly identifies the position of the image vortex. We then consider the case of a single vortex in a harmonic trapping potential, where SINDy suggests that the most parsimonious model is that of a single image vortex.  Finally, we consider more general trapping potentials of the form $V\propto|\rr|^{2p}$, and show that as $p$ increases the position of the image vortex tends towards the hard wall limit. The layout of this paper is as follows: Section~\ref{sec:pvm} introduces the point vortex model, with particular focus on its application to superfluids of ultra-cold atomic BECs. Section~\ref{sec:SINDy} contains an overview of the SINDy framework, followed by its application to a vortex in a hard wall trap. Section~\ref{sec:results} contains the main results of this paper, the SINDy framework is applied to a single vortex in a harmonic trap, and then trapping potentials with higher exponents are considered. Final conclusions are presented in Section~\ref{sec:conclusion}, along with potential avenues for future work. 

\section{The Point Vortex Model}
\label{sec:pvm}

In general, the point vortex model can be expressed as \cite{Newton}
\begin{equation}
   \frac{d}{dt}\rr_j = \vv_j 
   + \mathcal{D}\left(\rr_j,q_j,\mathcal{F}\right) + \mathcal{B}\left(\rr_j,q_j,\mathcal{F}\right),
    \label{eqn:general_pvm}
\end{equation}
where 
\begin{equation}
    \vv_j = \sum_{k\neq j} \frac{q_k}{r_{jk}^2} \left(\begin{matrix} - y_{jk} \\ x_{jk} \end{matrix} \right)
    \label{eqn:velocity}
\end{equation}
is the velocity of the $j$-th vortex induced by the other vortices in the system \cite{Fetter1966_IV} and $q_j$ is the charge of the $j$-th vortex.
We have written Eqn.~\eqref{eqn:general_pvm} in a general form, where $\mathcal{D}$ is the term due to the background density of the fluid, $\mathcal{B}$ is the term associated with any boundaries that the fluid possesses, and $\mathcal{F}$ is a set of properties of the fluid, including (but not limited to) the background density and the speed of sound in the fluid. 
It is possible to derived the exact equation of motion for a non-relativistic vortex filament moving in a non-homogeneous fluid background from a complex Ginzburg-Landau equation \cite{Tornkvist1997} (see also Ref.~\cite{Groszek2018}). The expression for the velocity field induced by local density variation is given by 
\begin{equation}
    \mathcal{D}\left(\vv_j,q_j,\mathcal{F}\right) = - q_j \boldsymbol{\hat{z}}\times \nabla \ln \rho \big|_{\rr_j}, 
\end{equation}
where $\rho\left(\rr\right)$ is the density of the fluid away from the vortex core, and the expression is evaluated at the position of the $j$-th vortex. 

For a hard wall trap with boundary $\partial\mathcal{B}$, the method of images can be used to ensure that there is no fluid flow across the boundary. This condition can be expressed as  
\begin{equation}
    \boldsymbol{J}\left(\partial\mathcal{B}\right)\boldsymbol{ \cdot n} = 0,
    \label{eqn:boundary_conditions}
\end{equation}
where $\boldsymbol{n}$ is a vector that is normal to the boundary, and we can write the mass-current of the fluid as $\boldsymbol{J}(\rr)=\rho(\rr)\vv(\rr)$, the product of the fluid density $\rho(\rr)$ and the fluid velocity $\vv(\rr)$. For a fluid with a well-defined boundary, this condition can be imposed by solving the associated Dirichlet problem for a Green's function of the second kind; for a vortex with charge $q_v$ at position $\rr_v$ within a circular trap of radius $R$, the image vortex would have charge $-q_v$ and will be located at $\rr_I = R^2\rr_v/|\rr_v|^2$, outside the trap \cite{Newton}. Typically one can take advantage of symmetries of the domain, and add an image vortex with appropriate charge at a location outside the domain to ensure that Eqn.~\eqref{eqn:boundary_conditions} is satisfied on the boundary. This will put a constraint on the velocity $\vv$ in the point vortex model. For a superfluid BEC that is confined in a  circularly symmetric potential of the form
\begin{equation}
    V(\rr) \propto |\rr|^{2p}, 
\end{equation}
where $p$ is positive, the density is well approximated by the Thomas-Fermi profile  \cite{Dalfovo1999}
\begin{equation}
    \rho_\mathrm{TF}(\rr) = \begin{cases}
        1 - V(\rr), \qquad \text{ when } \ 1-V(\rr)\geq 0, \\ 0 \qquad \qquad \qquad \text{otherwise.}
    \end{cases}
    \label{eqn:ThomasFermi}
\end{equation}
As a consequence, when $\rr$ approaches the edge of the trap, $\rho\left(\rr\to\partial\mathcal{B}\right)\to 0$ and hence  $\rho\vv \boldsymbol{\cdot n} \to 0$, trivially satisfying Eqn.~\eqref{eqn:boundary_conditions} without imposing any constraint on the velocity field.  The lack of constraint on the velocity has led to debate within the superfluid community as to the correct location for an image vortex in such a system. Indeed, previous works have argued that no image vortices are required because the fluid satisfies the boundary condition in Eqn.~\eqref{eqn:boundary_conditions} \cite{Kim2004}, or a single image vortex should be added, either with \cite{Middlekamp2011} or without \cite{Richaud2022} a factor to correct for the harmonic trapping potential. More recently, Ref.~\cite{Groszek2018} suggested that an infinite series of image vortices should be added; this captures the underlying physics of the system very well, although it is necessary to truncate this series for computational purposes. In the remainder of this paper we determine whether it is possible to use a single image vortex to capture the dynamics of a vortex in a range of trapping potentials.

\section{Sparse Identification of Nonlinear Dynamics (SINDy)}
\label{sec:SINDy}
\subsection{Overview}

The Sparse Identification of Nonlinear Dynamics (SINDy) framework is motivated by the assumption that the majority of physical models can be written in the form 
\begin{equation}
    \boldsymbol{\dot{x}} = \frac{d}{dt} \boldsymbol{x} = \boldsymbol{f} \left(\boldsymbol{x}\right),
    \label{eqn:general_dynamics}
\end{equation}
where the function $\boldsymbol{f}$ consists of relatively few terms \cite{Brunton2016}. The challenge is then to identify the function $\boldsymbol{f}$ from some observed values of $\boldsymbol{x}$. Let's suppose that the model of interest has $N$ variables, $\{x_1, x_2,\dots,x_N\}$, which can each be sampled on $T$ occasions, $\{t_1,t_2,\dots,t_T\}$, along with their derivatives (note that the derivatives may have to be numerically computed). These observations can be written as 
\begin{eqnarray}
    \boldsymbol{X} &=& \left[ \begin{matrix}
         x_1\left(t_1\right) & x_2\left(t_1\right) & \cdots & x_N\left(t_1\right) \\
         x_1\left(t_2\right) & x_2\left(t_2\right) & \cdots & x_N\left(t_2\right) \\
         \vdots & \vdots & \ddots & \vdots \\
         x_1\left(t_T\right) & x_2\left(t_T\right) & \cdots & x_N\left(t_T\right) \end{matrix}\right] , \nonumber \\
         \ & \ & \\
         \boldsymbol{\dot{X}} &=& \left[ \begin{matrix}
         \dot{x}_1\left(t_1\right) & \dot{x}_2\left(t_1\right) & \cdots & \dot{x}_N\left(t_1\right) \\
         \dot{x}_1\left(t_2\right) & \dot{x}_2\left(t_2\right) & \cdots & \dot{x}_N\left(t_2\right) \\
         \vdots & \vdots & \ddots & \vdots \\
         \dot{x}_1\left(t_T\right) & \dot{x}_2\left(t_T\right) & \cdots & \dot{x}_N\left(t_T\right) \end{matrix}\right], \nonumber          
\end{eqnarray}
where $\boldsymbol{X},\boldsymbol{\dot{X}}\in\mathbb{R}^{T\times N}$. It is then possible to compute a library of candidate functions, $\boldsymbol{\Theta}\left(\boldsymbol{X}\right)$, for the function $\boldsymbol{f}$. The library may consist of any function, but is typically informed by prior knowledge of the physical system that the model will describe. For a library of polynomial terms, the library will be written as
\begin{equation}
    \boldsymbol{\Theta}\left(\boldsymbol{X}\right) = \left[ \begin{matrix} \boldsymbol{1} & \boldsymbol{X} & \boldsymbol{X}^2 & \boldsymbol{X}^3 & \cdots \end{matrix} \right].
\end{equation}
Each of the $\boldsymbol{X}^n$ is a block matrix with $T$ rows, where each column containing one term of the multinomial expansion of the columns of $\boldsymbol{X}$, so the matrix $\boldsymbol{X}^n$ therefore has $\left(N+n-1\right)!/(N-1)!n!$ columns. The matrix of candidate functions $\boldsymbol{\Theta}\left(\boldsymbol{X}\right)\in\mathbb{R}^{T\times S}$. Using the candidate library, the system can be described by the linear equation
\begin{equation}
    \boldsymbol{\dot{X}} = \boldsymbol{\Theta}\left(\boldsymbol{X}\right) \boldsymbol{\Xi},
    \label{eqn:general_SINDy}
\end{equation}
where $\boldsymbol{\Xi}=\left[\begin{matrix} \boldsymbol{\xi}_1 & \boldsymbol{\xi}_2 & \cdots & \boldsymbol{\xi}_N \end{matrix}\right]$ are the coefficients of the model, $\boldsymbol{\Xi}\in\mathbb{R}^{S\times N}$. If $T>S$ (i.e., there are more samples of the data than there are terms in the library of candidate functions)  Eqn.~\eqref{eqn:general_SINDy} is an over-determined system, which allows us to find coefficient vectors $\boldsymbol{\xi}$ that are sparse. This can be expressed as least-squares regression with a penalty for the number of terms, 
\begin{equation}
    \hat{\boldsymbol{\xi}}_k = \text{argmin}_{\boldsymbol{\xi}_k} \| \boldsymbol{\dot{X}}_{k,:} - \boldsymbol{\Theta}\left(\boldsymbol{X}\right)\boldsymbol{\xi}_k\|_2^2 + \lambda \|\boldsymbol{\xi}_k\|_0,
    \label{eqn:SINDY_formal}
\end{equation}
where $\boldsymbol{\dot{X}}_{k,:}$ is the $k^\mathrm{th}$ row of $\boldsymbol{\dot{X}}$ \cite{Fasel2021}. This penalty provides flexibility when constructing the library of candidate functions as it will exclude irrelevant terms; it also has the advantage of preventing the over-fitting of models. The optimization in Eqn.~\eqref{eqn:SINDY_formal} can be solved using sequential thresholded least squares where $\lambda$ controls the sparsification of the system \cite{Brunton2016}. In the language of sparse regression, we can consider the terms in Eqn.~\eqref{eqn:general_SINDy} to be the response (the derivative, $\boldsymbol{\dot{X}}$), the features (the library terms $\boldsymbol{\Theta}(\boldsymbol{X})$), and the loadings (the coefficients, $\boldsymbol{\Xi}$), respectively. It is worth noting that the penalty term, $\|\boldsymbol{\xi}_k\|_0$, in Eqn.~\eqref{eqn:SINDY_formal} is not unique - it is possible to use other regularisation terms such as least absolute shrinkage and selection operator (LASSO) regression \cite{Tibshirani1996} and sparse relaxed regularisation regression (SR3) \cite{Zheng2018}.

\subsection{SINDy on Rational Functions}
\label{subsec:rational_SINDy}
As well as considering methods for learning dynamical equations of the form Eqn.~\eqref{eqn:general_dynamics}, it is necessary to consider systems that may contain rational functions, where for some $k\in\{1,\dots,N\}$ 
\begin{equation}
     \dot{x}_k = \frac{f(\boldsymbol{x})}{g(\boldsymbol{x})},
    \label{eqn:rational_dynamics}
\end{equation}
and it's necessary to find both $f$ and $g$. A variation of the SINDy algorithm, \emph{implicit}-SINDy, developed in Ref.~\cite{Mangan2016} allows estimations to be made for the functions $f$ and $g$, which would otherwise be computationally unwieldily (the size of the library would be impractical if the terms contained multiple possible combinations of ratios of polynomials). The starting point is to multiply by $g(\boldsymbol{x})$ to write the implicit equation  
\begin{equation}
    f(\boldsymbol{x}) - \dot{x}_k g(\boldsymbol{x}) = 0.
\end{equation}
This means that for each variable $x_k$, the library of candidate functions can be written as
\begin{equation}
\boldsymbol{\Theta}\left(\boldsymbol{X}\right) = \left[ \begin{matrix} \boldsymbol{\Theta}_f\left(\boldsymbol{X}\right) & \boldsymbol{\dot{X}}_k \boldsymbol{\cdot \Theta}_g \left(\boldsymbol{X}\right) \end{matrix}\right],
\end{equation}
where $\boldsymbol{\Theta}_f$ and $\boldsymbol{\Theta}_g$ are the libraries of the candidate terms for $f$ and $g$ respectively, and we use $\boldsymbol{\dot{X}}_k \boldsymbol{\cdot \Theta}_g \left(\boldsymbol{X}\right)$  to mean that every column in $\boldsymbol{\Theta}_g$ is element-wise multiplied by the vector $\boldsymbol{\dot{X}}_k$. It then remains to find the sparse set of coefficients $\boldsymbol{\xi}_k$ such that
\begin{equation}
    \boldsymbol{\Theta} \left(\boldsymbol{X}_k\right) \boldsymbol{\xi}_k = \boldsymbol{0}.
    \label{eqn:SINDy_null}
\end{equation}
In this case, using the original sparse regression algorithm in Ref.~\cite{Brunton2016} would return a zero vector. Instead, as advocated by Ref.~\cite{Mangan2016}, the best approach is to identify vectors in the null space of $\boldsymbol{\Theta}\left(\boldsymbol{X}_k\right)$, and find a linear combination of these vectors that gives the sparsest resulting vector. For a given $k$, the null space of the matrix $\boldsymbol{\Theta}\left(\boldsymbol{X}_k\right)$ can be written as 
\begin{equation}
    \mathcal{N}_k = \{ \boldsymbol{\zeta}_1, \boldsymbol{\zeta}_2, \dots, \boldsymbol{\zeta}_\eta \}. 
\end{equation}
It is then possible to use the alternating directions method, Ref.~\cite{Qu2015}, to find the coefficients $\boldsymbol{\alpha}=\{\alpha_1,\alpha_2,\dots,\alpha_\eta\}$ such that 
\begin{equation}
    \boldsymbol{\xi}_k = \alpha_1\boldsymbol{\zeta}_1 + \alpha_2\boldsymbol{\zeta}_2 + \cdots + \alpha_\eta \boldsymbol{\zeta}_\eta
\end{equation}
is sparse. This process needs to be repeated for each variable $x_k$. Once the coefficients have been recovered, they can then be thresholded: the simplest way to do this is to take a value of the sparsification parameter $\lambda$, and then set
\begin{equation}
    \left[\boldsymbol{\xi}_k^*\right]_j = \begin{cases}
        \left[\boldsymbol{\xi}_k\right]_j \qquad \text{ if } \  \left|\left[\boldsymbol{\xi}_k\right]_j\right| > \lambda, \\
        \ \\ 0 \qquad \text{otherwise,}
    \end{cases} \qquad \text{ for } \ j\in\{1,\dots,N\}.
\end{equation}
The resulting equations of motion can then be recovered from the thresholded vector of coefficients $\boldsymbol{\xi}_k^*$. We will discuss the value of the sparsification parameter $\lambda$ in the following sections. 

\subsection{Synthetic Data: Hard Wall Trap}
\label{subsec:synthetic}
In this section, synthetic data is generated for a single point vortex in a hard wall trap, where it is known that the vortex will precess around the trap at a constant speed at a fixed distance from the trap center. The functional form of the equations of motion are known in this case \cite{Newton}, but it provides an illustrative example for implementing SINDy on rational functions, as discussed in the previous subsection.

For a hard wall trap of radius $R$, it is known that a vortex with charge $q_v$ at position $\left(x_v,y_v\right)$ will have an image vortex at 
\begin{equation}
    \left(\begin{matrix} x_I \\ y_I \end{matrix}\right) = \frac{R^2}{x_v^2 + y_v^2}\left(\begin{matrix} x_v \\ y_v \end{matrix} \right),
    \label{eqn:hw_image_vortex}
\end{equation}
with charge $-q_v$. In the case of a single vortex, the point vortex model in Eqn.~\eqref{eqn:general_pvm} reduces to 
\begin{equation}
    \frac{d}{dt}\left( \begin{matrix} x_v \\ y_v \end{matrix} \right) = \frac{-q_v}{x_v^2 + y_v^2 - R^2} \left(\begin{matrix} -y_v \\ x_v \end{matrix} \right),
    \label{eqn:Hard_Wall}
\end{equation}
where it may be assumed that the vortex is far enough from the boundary that variations in the background density are negligible. To generate the synthetic data, we evolve Eqn.~\eqref{eqn:Hard_Wall} with $q_v=1$ and the initial condition $\left(x_v,y_v\right)=\left(0,0.7R\right)$ in a trap with radius $R=32$ up to $t=2.5\times10^4$ with a timestep of $dt=10^{-1}$. The time derivatives, $\dot{x}_v$ and $\dot{y}_v$, can then be approximated using a simple finite difference scheme. 

Having generated the synthetic data, we apply the SINDy algorithm for rational functions to see if it can correctly identify the point vortex model. Writing Eqn.~\eqref{eqn:Hard_Wall} in implicit form yields
\begin{equation}
    \begin{matrix}
        q_v y_v + \dot{x}_v \left(R^2 - x_v^2 - y_v^2\right) & = 0, \\
        q_v x_v - \dot{y}_v \left(R^2 - x_v^2 - y_v^2\right) & = 0.
    \end{matrix}
\end{equation}
This suggests that the library for the numerator and the denominator of the rational function should include terms up to and including quadratic terms, i.e., $ \left[ \begin{matrix}
        1 & x_v & y_v & x_v^2 & x_vy_v & y_v^2
    \end{matrix}\right] $. The result is a library for each of the $\dot{x}$ and $\dot{y}$ variables which is
    \begin{widetext}
\begin{equation}
\begin{matrix}
    \boldsymbol{\Theta}_x & = & [ 1 \quad x_v \quad y_v \quad x_v^2 \quad x_vy_v \quad y_v^2 \quad \dot{x}_v \quad \dot{x}_v x_v \quad \dot{x}_v y_v \quad \dot{x}_v x_v^2 \quad \dot{x}_v x_v y_v \quad \dot{x}_v y_v^2 ], \\
    \boldsymbol{\Theta}_y  & = & [ 1 \quad x_v \quad y_v \quad x_v^2 \quad x_vy_v \quad y_v^2 \quad \dot{y}_v \quad \dot{y}_v x_v \quad \dot{y}_v y_v \quad \dot{y}_v x_v^2 \quad \dot{y}_v x_v y_v \quad \dot{y}_v y_v^2 ] .
\end{matrix}
\label{eqn:synthetic_library}
\end{equation}
\end{widetext}
We then follow the algorithm described in the previous section, finding the null spaces, $\mathcal{N}_x$ and $\mathcal{N}_y$, of the matrices $\boldsymbol{\Theta}_x$ and $\boldsymbol{\Theta}_y$, and using the alternating directions method to obtain sparse vectors $\boldsymbol{\xi}_x$ and $\boldsymbol{\xi}_y$.

\begin{figure*}
    \centering
    \includegraphics[scale=1.0]{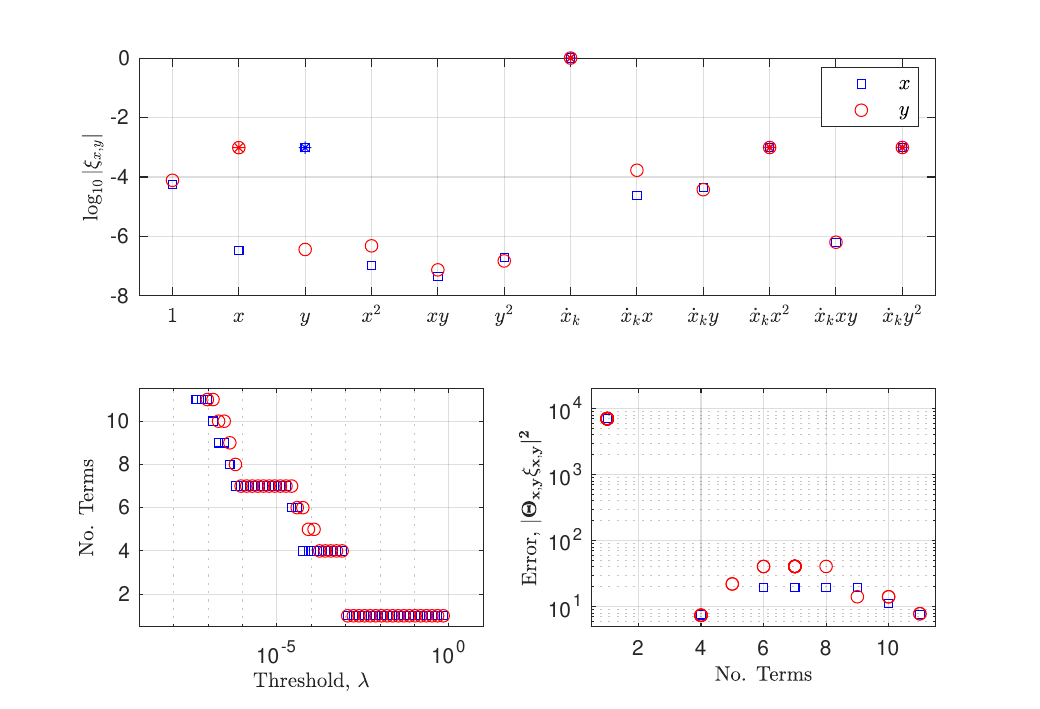}
    \caption{Applying the SINDy algorithm to synthetic data for a hard walled trap. Top panel: the absolute value of the returned coefficients for the $x$ (blue squares) and $y$ (red circles) equations of motion. The correct coefficients from Eqn.~\eqref{eqn:Hard_Wall} are indicated with asterisks. Bottom left: increasing the thresholding parameter $\lambda$ reduces the number of non-zero terms in the coefficient vector. Bottom right: the model error is calculated, the lowest error combined with the most parsimonious model gives the best choice of model coefficients.}
    \label{fig:synthetic_results}
\end{figure*}

The results of this computation on the synthetic data is in Fig.~\ref{fig:synthetic_results}. Since it is difficult to know the value of the sparsification parameter in advance, we sequentially increase the value of $\lambda$ from $10^{-8}$ to $10^0$, and record the number of non-zero terms after removing values of $|\boldsymbol{\xi}_{x,y}|$ that are smaller than the sparsification parameter. As the bottom left panel of Fig.~\ref{fig:synthetic_results} shows, increasing the value of $\lambda$ decreases the number of non-zero coefficients. When choosing a value of the sparsification parameter, the goal is to balance model accuracy (the smallest value of $\|\boldsymbol{\Theta}_{x,y}\boldsymbol{\xi}_{x,y}\|_2^2$) against the parsimony of the model (the fewest terms). As shown in the bottom right panel of Fig.~\ref{fig:synthetic_results}, in the case of the hard wall trap a model with 4 terms gives the lowest error (the error from the model with 4 terms is slightly lower than the error from the model with 12 terms), and is clearly more sensible than a model with 12 terms that is almost certainly the result of over-fitting the data. 

\begin{figure*}
    \centering
    \includegraphics[scale=1.0]{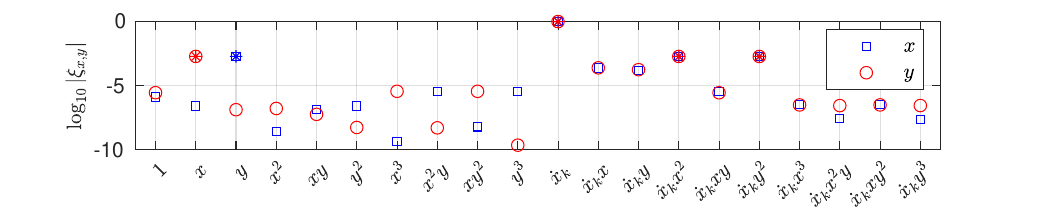}
    \caption{The absolute value of the returned value of the coefficients for the $x$ (blue squares) and $y$ equations of motion of a single vortex in a hard wall trap. These coefficients are for the extended library of candidate functions given in Eqn.~\eqref{eqn:synthetic_extended_library}, which now include cubic terms. The correct terms in the model, indicated by asterisks, are identified by the SINDy algorithm even with this extended library}.
    \label{fig:synthetic_cubic}
\end{figure*}

Since the SINDy algorithm has correctly identified the number of terms in the model, we turn our attention to the returned coefficients. The returned coefficients lead to the equations of motion 
\begin{equation}
    \begin{matrix}
        0.9919 y_v + \dot{x}_v \left(1023.9985 - 1.0084x_v^2 - 1.0061y_v^2\right) & = 0, \\
        0.9935 x_v - \dot{y}_v \left(1023.9985 - 1.0046x_v^2 - 1.0046y_v^2\right) & = 0. 
    \end{matrix}
\end{equation}
Note that each of these equations have been multiplied by $R^2$. These values all have a relative error that is less than $0.01\%$. In order to check that the results of applying the SINDy algorithm to the synthetic data isn't an artifact of the library that was chosen, we extend the libraries in Eqn.~\eqref{eqn:synthetic_library} to include cubic terms, so
\begin{widetext}
\begin{equation}
    \boldsymbol{\Theta}_x = \left[ 1 \quad x_v \quad \dots \quad x_v^3 \quad x_v^2 y_v \quad x_v y_v^2 \quad y_v^3 \quad \dot{x}_v \quad \dots \quad \dot{x}_v x_v^3 \quad \dots \quad \dot{x}_v y_v^3 \right]
    \label{eqn:synthetic_extended_library}
\end{equation}
\end{widetext}
and likewise for $\boldsymbol{\Theta}_y$. The coefficients returned for these candidate models are shown in Fig.~\ref{fig:synthetic_cubic}. The SINDy algorithm still identifies the correct terms for the model, however we note that in this case the relative error in the values of the coefficients has crept up to roughly $1\%$. This increase in error may be due to the increase in multi-collinearity between terms in the extended library.

We end this section by commenting on the data requirements of the SINDy algorithm in order to correctly identify the point vortex model. While the results presented in this section are for a particular value of $R$ and initial conditions $\left(x_v(0),y_v(0)\right)$, we've confirmed that SINDy correctly identifies model terms as these parameters are varied. Of particular importance, however, is the sampling duration of the system (for a full discussion, see Ref.~\cite{Champion2019}) where we found that SINDy required the synthetic data to be such that the vortex had completed at least 6 full orbits of the trap. In addition, we found that the accuracy of the identified coefficients improved as the initial vortex position was closer to the trap wall. This is unsurprising, as the velocity of a vortex decays proportionally to $1/r^2$, and so the influence of the image vortex is stronger as the vortex is closer to the edge of the trap.

\section{Gross-Pitaevskii Results}
\label{sec:results}

\subsection{Data}
To obtain the data for this section, we evolve the dimensionless 2D Gross-Pitaevskii Equation (GPE)
\begin{equation}
    i \frac{\partial \psi}{\partial t} = \left[-\frac{1}{2}\left(\frac{\partial^2}{\partial x^2} + \frac{\partial^2}{\partial y^2}\right) + \left(\frac{r}{R}\right)^{2p} + g|\psi|^2 - 1 \right] \psi,
    \label{eqn:2D_GPE}
\end{equation}
using an adaptive Runge-Kutta 45 scheme with an error tolerance of $10^{-8}$ \cite{XMDS2}. This equation accurately describes an ultra-cold dilute atomic Bose-Einstein Condensate (BEC) in the zero-temperature limit, with a macroscopic wavefunction $\psi$ \cite{Primer} (further details of the non-dimensionalisation of Eqn.~\eqref{eqn:2D_GPE} can be found in Appendix~\ref{appendix:GPE}). We initialise the system by making the substitution $t\to it$ and evolving the system while reimposing the phase of a vortex with charge $q_v=1$ at $\left(x_0,y_0\right)$ after each time step. This ``imaginary time'' evolution leads to the ground state of the system, removing transient sound waves that may distort the trajectory of the vortex, and negating the requirement to know the density profile of the vortex core. 

As discussed in the previous section, when trying to identify the position of an image vortex, it is advantageous to choose an initial vortex location where the effects of the image vortex will dominate. As such, we take the characteristic width of the trapping potential to be $R=32$ (which is easily realisable in an experiment), and we set $\left(x_0,y_0\right)=(0,0.7R)$. We have confirmed that the model identified by the SINDy framework is robust against varying the initial vortex position. After each GPE simulation, we extract the position of the vortex using a plaquette method that is standard in GPE vortex dynamic literature. The position obtained has an uncertainty $\Delta x$ in the $x$ direction ($\Delta y$ in the $y$ direction) that corresponds to the grid spacing of the GPE simulation. As such - we apply a small amount of Gaussian smoothing to the trajectory and then compute the derivatives using a five-point stencil.

\subsection{Model Identification in a Harmonic Trap}
We now consider the case of a single vortex in a harmonic trapping potential, which we write as $V(\rr)\propto|\rr|^2$. This form of potential is commonplace in BEC experiments that take advantage of optical and magnetic trapping potentials \cite{Gorlitz2001}, and is therefore of great interest for experimental vortex dynamics. 

We begin by generalising the position of the image vortex given in Eqn.~\eqref{eqn:hw_image_vortex}. We suppose the position of the image vortex to be $\rr_I = \varphi^2 \rr_v/ |\rr_v|^2$, for an unknown distance $\varphi$, an approximation that is later confirmed by the SINDy analysis. The evolution of the vortex described by Eqn.~\eqref{eqn:general_pvm} will be given by 
\begin{equation}
    \frac{d}{dt} \left( \begin{matrix} x_v \\ y_v \end{matrix}\right) = \frac{-q_v}{x_v^2 + y_v^2 - \varphi^2}\left( \begin{matrix} - y_v \\ x_v \end{matrix} \right) - q_v \left(\begin{matrix} - w_y\left(\rr_v\right) \\ \quad w_x\left(\rr_v\right) \end{matrix} \right),
    \label{eqn:power_trapped_vortex}
\end{equation}
where 
$\left(w_x\left(\rr_v\right),w_y\left(\rr_v\right)\right)=\nabla\ln\rho_\mathrm{TF}|_{\rr_v}$ are the terms due to variation in the Thomas-Fermi density \cite{Tornkvist1997}. This allows us to write Eqn.~\eqref{eqn:power_trapped_vortex} in the implicit form
    \begin{equation}
    \begin{matrix}
        q_v y_v + \left( \dot{x}_v - q_v w_y\right) \left(\varphi^2 - x_v^2 - y_v^2\right) & = 0, \\
        q_v x_v - \left(\dot{y}_v + q_v w_x \right) \left(\varphi^2 - x_v^2 - y_v^2\right) & = 0.
        \label{eqn:point_vortex_fit}
    \end{matrix}
\end{equation}
For a harmonic trap (for the general expression see Eqn.~\eqref{eqn:background_terms}) the background density terms can be written as $\left(w_x,w_y\right)= 2\rr_v/\left( |\rr_v|^2 - R^2\right)$. These terms can be computed from the vortex positions, allowing us to write the implicit equation
\begin{equation}
    \begin{matrix}
        q_v y_v + \dot{X}_v\left(\varphi^2 - x_v^2 - y_v^2\right) & =0, \\
        q_v x_v - \dot{Y}_v\left(\varphi^2 - x_v^2 - y_v^2\right) &=0,
    \end{matrix}
    \label{eqn:harmonic_eom}
\end{equation}
where $\dot{X}_v=\dot{x}_v - q_v w_y$ and $\dot{Y}_v=\dot{y}_v+q_v w_x$. 

\begin{figure*}
    \centering
    \includegraphics[scale=1.0]{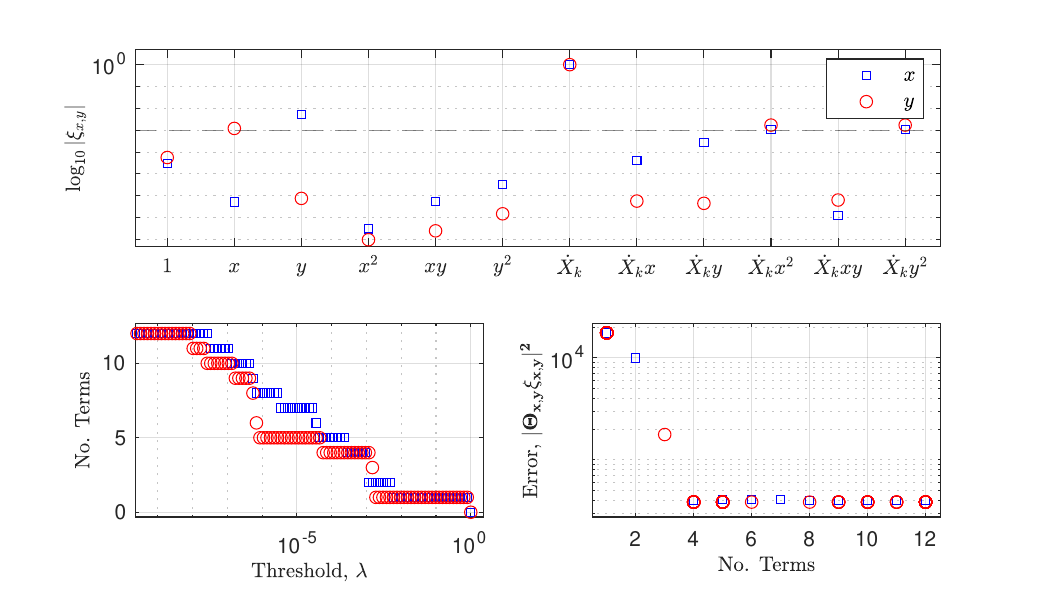}
    \caption{Applying the SINDy algorithm to GPE data of a single vortex in a harmonic trap. Top panel: the absolute value of the returned coefficients for the $x$ (blue squares) and $y$ (red circles) equations of motion; the dashed line $R^{-2}$ is added as a guide to the eye. Bottom left: increasing the thresholding parameter reduces the number of non-zero terms. Bottom right: the error is calculated for models that contain the given number of terms.}
    \label{fig:harmonic_model}
\end{figure*}

The form of Eqn.~\eqref{eqn:harmonic_eom} suggests  a library of the form
\begin{widetext}
\begin{eqnarray}
    \boldsymbol{\Theta}_x & = & \left[ 1 \quad x_v \quad y_v \quad x_v^2 \quad x_vy_v \quad y_v^2 \quad \dot{X}_v \quad \dot{X}_v x_v \quad \dot{X}_v y_v \quad \dot{X}_v x_v^2 \quad \dot{X}_v x_v y_v \quad \dot{X}_v y_v^2 \right], \nonumber \\
    \ & \ & \label{eqn:harmonic_library} \\
    \boldsymbol{\Theta}_y  & = & \left[ 1 \quad x_v \quad y_v \quad x_v^2 \quad x_vy_v \quad y_v^2   \quad \dot{Y}_v \ \quad \dot{Y}_v x_v \ \quad \dot{Y}_v y_v \ \quad \dot{Y}_v x_v^2 \ \quad \dot{Y}_v x_v y_v \ \quad \dot{Y}_v y_v^2 \right], \nonumber 
\end{eqnarray}
\end{widetext}
for the SINDy framework. The results of this analysis can be found in Fig.~\ref{fig:harmonic_model}. Unsurprisingly, the largest coefficients are for the $\dot{X}_v$ and $\dot{Y}_v$ terms in each equation of motion -- from Eqn.~\eqref{eqn:harmonic_eom} we expect that these are $\varphi^2$, where $\varphi\sim \mathcal{O}(R)$. As in the case of the synthetic data, the optimal sparsification value of the sparsification parameter is unknown \emph{a priori}, so we sequentially increase the value of $\lambda$ from $10^{-10}$ to $10^0$, in order to identify the most parsimonious model. The sharp reduction in error between models with 3 terms and models with 4 terms that can be observed in the bottom right panel of Fig.~\ref{fig:harmonic_model} suggests that a model with 4 terms is the most parsimonious.

\begin{figure*}
    \centering
    \includegraphics[scale=1.0]{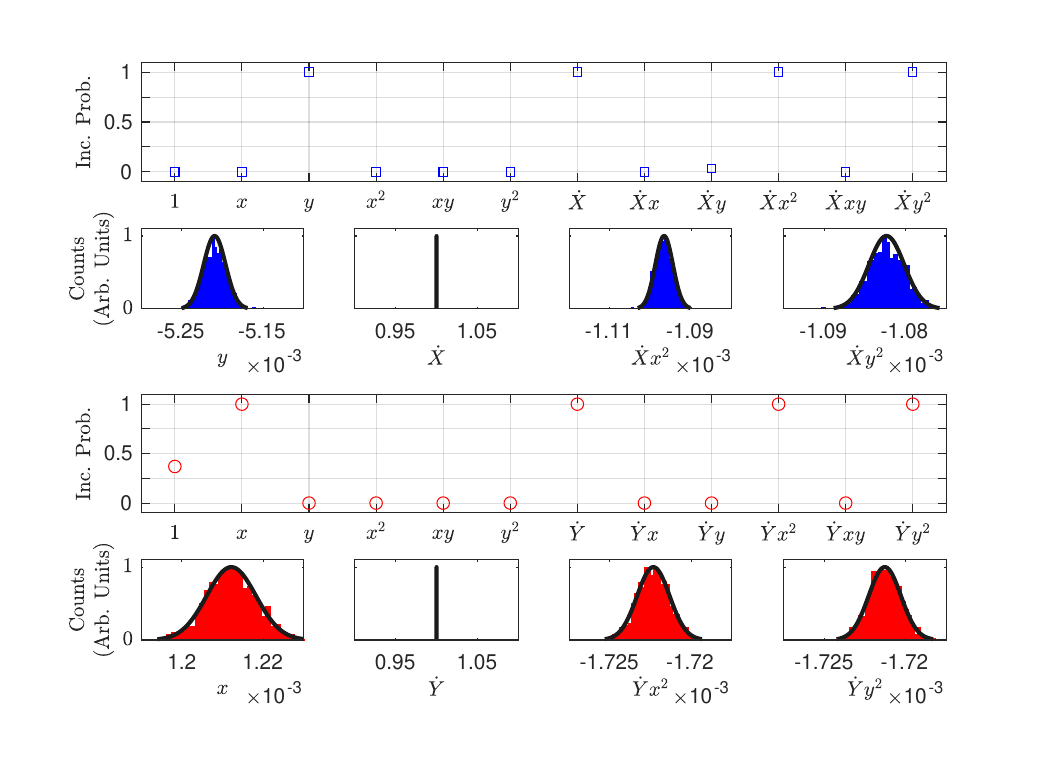}
    \caption{Application of Ensemble SINDy to the data obtained from the GPE with harmonic trapping. First (third) row: the inclusion probability of terms in the $\dot{X}$ ($\dot{Y}$) equations of motion. Second (fourth) row: normalised histogram and probability density function of the fitted normal distribution for the value of the coefficients of the terms $y$, $\dot{X}$, $\dot{X}x^2$ and $\dot{X}y^2$ ($x$, $\dot{Y}$, $\dot{Y}x^2$ and $\dot{Y}y^2$).}
    \label{fig:ensemble_harmonic}
\end{figure*}

In order to obtain the values of the coefficients in Eqn.~\eqref{eqn:harmonic_eom}, we now apply the ensemble-SINDy (E-SINDy) algorithm \cite{Fasel2022} to the data. This algorithm takes advantage of two well-known statistical techniques: bootstrap aggregation \cite{Briemann1996} and bagging \cite{Buhlmann2012}. As has previously been discussed, methods for data-driven model discovery are sensitive to noise \cite{Kaiser2018,Reinbold2020,Fasel2022} and this extends to implicit-SINDy calculations where noise in the data will propagate into derivatives of the data, and may be exacerbated in the $\boldsymbol{\dot{X}}_k\boldsymbol{\cdot\Theta}_g\left(\boldsymbol{X}\right)$ terms of the library. E-SINDy (which we now take to include performing ensemble techniques on implicit SINDy calculations) provides both inclusion probabilities and uncertainty estimates on the values of the coefficients identified by the SINDy framework \cite{Fasel2022}.

The data obtained from the GPE simulations of a single vortex in a harmonic trap consists of observations of the vortex coordinates $\left(x_v(t_i),y_v(t_i)\right)$ and their numerically obtained derivatives $\left(\dot{x}_v(t_i),\dot{y}_v(t_i)\right)$ for $i=1,\dots,T$, where $T\approx10^5$. To perform E-SINDy on this data, we sample $T$ observations from this data set with replacement, and perform the SINDy algorithm described in Sec.~\ref{sec:SINDy} on the sample dataset. We perform this $N_{bag}$ times, saving the obtained values of the coefficients $\xi_k$. After applying SINDy to each sample, we threshold the coefficients: if $|\xi_k|<\lambda$, we set $\xi_k=0$. Doing so allows us to calculate the inclusion probability,
\begin{equation}
    \Pr\left(\xi_k \text{ included in model}\right) = \frac{1}{N_{bag}}\sum_{j=1}^{N_{bag}} c_{jk},
\end{equation}
where $c_{jk}=1$ if $|\xi_k|>\lambda$ on the $j$-th sampling and is zero otherwise. Based on the results of Fig.~\ref{fig:harmonic_model}, we take $\lambda\approx 1/ \left(3R^2\right)$; this is sufficiently small that it will exclude the majority of terms, but not so small that it will exclude ``near miss'' terms that may be included in the model. The results of our E-SINDy calculation with $N_{bag}=10^3$ are in Fig.~\ref{fig:ensemble_harmonic}. The inclusion probability for both the $\dot{X}$ and $\dot{Y}$ equations of motion are consistent with the model in Eqn.~\eqref{eqn:harmonic_eom}, as we see that the expected terms are included with near 100\% probability. The coefficients of the $\dot{X}y$ term and the constant term in the $\dot{Y}$ equation of motion have non-zero inclusion probabilities, we suggest that this is due simply to noise in the data. In any case, the inclusion probabilities of these terms are relatively small. We are able to fit a normal distribution to the ensemble of coefficients - this gives us an approximation for the mean of the ensemble $\mu\left(\xi_k\right)$, and allows us to infer the uncertainty in this value, $\sigma\left(\xi_k\right)$, which we use in our approximation of $\varphi^2$, which we discuss in the following subsection. 

\subsection{Image Vortex in a General Trapping Potential}
We have used the SINDy framework to confirm that the case of a single vortex in both a harmonic and a hard-wall trap is well approximated by Eqn.~\eqref{eqn:harmonic_eom}. We now turn our attention to identifying the parameter $\varphi$ for a trap that has the form 
\begin{equation}
    V(x,y) = \left(\frac{x^2 + y^2}{R^2}\right)^p,
    \label{eqn:general_trap}
\end{equation}
for some positive number $p$. In order to do this, we perform the E-SINDy algorithm described above to GPE simulations of a single vortex in a power-law trap for different values of $p$, now taking $N_{bag}=10^4$. 

Applying the E-SINDy algorithm for each value of $p$, we find the same terms are identified as in the case of a harmonic trap, suggesting that the most parsimonious model has the same functional form as Eqn.~\eqref{eqn:power_trapped_vortex}, which can be written implicitly as
\begin{equation*}
    \begin{matrix}
        q_v y_v + \dot{X}_v\left(\varphi^2 - x_v^2 - y_v^2\right) & =0, \\
        q_v x_v - \dot{Y}_v\left(\varphi^2 - x_v^2 - y_v^2\right) &=0. 
    \end{matrix}
\end{equation*}
As before, $\dot{X}_v=\dot{x}_v - q_v w_y$ and $\dot{Y}_v=\dot{y}_v+q_v w_x$, while for a trapping potential of the form given by Eqn.~\eqref{eqn:general_trap}, the background density terms are
\begin{equation}
\left(\begin{matrix} w_x\left(x,y\right) \\ w_y\left(x,y\right) \end{matrix}\right) = \frac{-2p \left(x^2 + y^2\right)^{p-1}}{R^{2p}-\left(x^2 + y^2\right)^p} \left(\begin{matrix} x \\ y \end{matrix}\right),
    \label{eqn:background_terms}
\end{equation}
see Appendix~\ref{appendix:image_vortex} for details. We proceed by simulating a single vortex in a number of trapping potentials using the Gross-Pitaevskii Equation, then extracting the vortex position and velocity. 

As previously discussed, the E-SINDy algorithm provides uncertainty estimates for each of the values of the coefficients in the identified model \cite{Fasel2022}. In our case, for each of the coefficients $\xi_k$ included in the model we fit a normal distribution with mean $\mu\left(\xi_k\right)$ and variance $\sigma^2\left(\xi_k\right)$, for a given $p$. When written in implicit form, Eqn.~\eqref{eqn:harmonic_eom}, we expect the majority of terms in the coupled equations of motion to be unity, so we can form six estimates of the parameter $\varphi^2$; these are given by the ratios $\mu\left(\xi_{j}\right)/\mu\left(\xi_k\right)$ where the pairs $(j,k)$ are given by \begin{widetext}
\begin{equation*}
    \left(j,k\right)\in\left\{\left(\dot{X}_v,y_v\right), \ \left(\dot{X}_v,\dot{X}_vx_v^2\right), \ \left(\dot{X}_v,\dot{X}_vy_v^2\right), \ \left(\dot{Y}_v,x_v\right), \ \left(\dot{Y}_v,\dot{Y}_vx_v^2\right), \ \left(\dot{Y}_v,\dot{Y}_vy_v^2\right)\right\}. 
\end{equation*}
\end{widetext} 
For each of the 6 methods for approximating $\varphi^2$, we can also estimate the variance as \cite{Diaz-Frances2013}
\begin{equation}
    \sigma^2\left( \frac{\xi_j}{\xi_k}\right) \approx \left[\frac{\mu\left(\xi_j\right)}{\mu\left(\xi_k\right)}\right]^2 \left[ \frac{\sigma^2\left(\xi_j\right)}{\mu\left(\xi_j\right)^2} + \frac{\sigma^2\left(\xi_k\right)}{\mu\left(\xi_k\right)^2} \right].
\end{equation}
The final mean and variance can be found by taking suitable linear combinations of $\mu\left(\xi_j\right)/\mu\left(\xi_k\right)$ and $\sigma^2\left(\xi_j/\xi_k\right)$, which give us an estimate and an uncertainty on the parameter $\varphi^2$ for each value of $p$. 

\begin{figure*}
    \centering
    \includegraphics[scale=1.0]{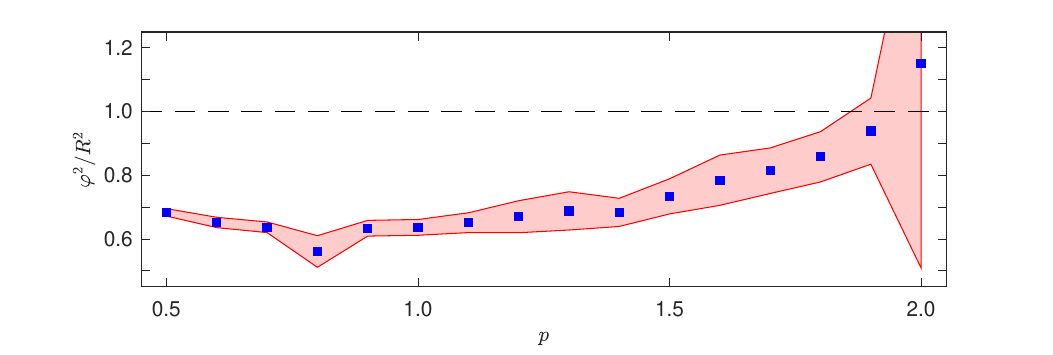}
    \caption{The position of the image vortex as a function of the trap power, $p$, can be estimated from the parameter $\varphi^2$. Blue squares indicate this for each of the values of $p$ considered, while the red shaded area indicates one standard deviation from the mean. The dashed black line indicates the hard wall limit, $\varphi^2\to R^2$.}
    \label{fig:varphi}
\end{figure*}

The results of this analysis can be found in Fig.~\ref{fig:varphi}. We see that as the power of the trap increases, the distance between the vortex and its image decreases; this will lead to a higher velocity (which is consistent with our numerical results) and is in agreement with the prediction of precession frequency in Ref.~\cite{Richaud2022}. As the power, $p$, increases the position of the image vortex smoothly increases and tends towards the known case of a hard-wall trap ($p\to\infty$). We also observe that the size of the uncertainty $\sigma^2\left(\varphi^2\right)$ increases as the power of the trapping potential increases -- this is most probably due to the fact that slower vortices completed fewer orbits of the trapping potential in these simulations, affecting the sampling available to the (E-)SINDy algorithm. 

\subsection{Comparison with Experiments}
We now compare our prediction of the point vortex model for the harmonic trap to the experiment of Freilich \emph{et al.}~\cite{Freilich2010}, who observed a single vortex line in a harmonically trapped quasi-2D condensate. We do this by taking $\varphi^2/R^2=0.6366$, and simulating a single vortex with charge $q_v=-1$ according to Eqn.~\eqref{eqn:power_trapped_vortex}. To enable comparison between experimental and numerical results, the simulation time $t_\mathrm{Sim}$ is multiplied by $655\mathrm{ms}\times\tau_\mathrm{E}dt/\left(\omega_r\tau_\mathrm{S}T_\mathrm{S}\right)$, where $\tau_\mathrm{E}=2\pi\times3.67\mathrm{Hz}$ was the observed precession frequency of the vortex line in the experiment and $\omega_r=2\pi\times 35.8\mathrm{Hz}$ was the radial trapping frequency \cite{Freilich2010}; the three dimensionless quantities, $1/\tau_\mathrm{S}$, $dt$ and $T_\mathrm{S}$, are the vortex period extracted from the simulation, the time step size in the simulation and the final integrator time, respectively.

\begin{figure*}
    \centering
    \includegraphics[scale=1.0]{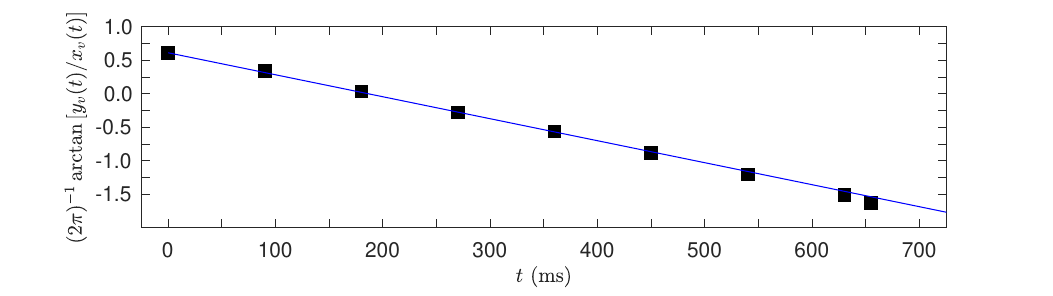}
    \caption{Comparison between the point vortex model in a harmonic trap and the experimental results of Ref.~\cite{Freilich2010}. Black squares show the experimental vortex phases after ballistic expansion. Blue curve shows the vortex phase given by the point vortex model in Eqn.~\eqref{eqn:power_trapped_vortex}.}
    \label{fig:experimental_comparison}
\end{figure*}

The comparison between the simulation and the experiment is shown in Fig.~\ref{fig:experimental_comparison}. As we can see, there is an excellent agreement between the experimental results and the point vortex simulation, indicating that the E-SINDy algorithm has correctly identified a parsimonious model for this system. We note that there is a small discrepancy at later times -- we suggest that this is due to the effect of dissipation, a small amount of which is present in the experimental system, but is not included in our model. The presence of dissipation in the system will cause the vortex to spiral as it precesses around the trapping potential. It remains an interesting extension to the model developed in this paper to consider the effects of dissipation, as such a model could then be used to infer the affect of vortex-sound interactions in the trap.

\section{Conclusion}
\label{sec:conclusion}

The point vortex model is a simple mathematical model for vortex dynamics, which is excellent for describing vortices in superfluids, such as Bose-Einstein Condensates (BECs), due to their quantized circulation and fixed core size. The model can be adapted to describe systems that have a hard-wall boundary, by adding an ``image vortex'' to the system. The exact nature of an image vortex in a harmonic trapping potential has been the subject of a number of debates, which is problematic given that BECs are typically trapped in a harmonic potential. This is mainly due to the fact that, aside from in hard-wall traps, the density of the fluid gradually approaches zero so there is no constraint on the velocity of the fluid at the edge of the trap.

In this work, we have applied the Sparse Identification of Non-linear Dynamics (SINDy) framework to a single vortex in a trapped BEC. The advantage of the SINDy framework is that it promotes sparsity in the model --- leading to a simple governing equation and avoiding over-fitting. We have shown that that we can approximate an image vortex in a circular harmonic trap using a single image vortex, placed at the correct distance from the trap centre. We have also shown that this approximation works for a variety of power law trapping potentials, and we have provided estimates for the uncertainty of the position of the image vortex in each case. We have shown that the position of this image vortex tends towards the hard-wall limit as the edge of the trapping potential becomes sharper. We have also shown that the model for a vortex in a harmonic trap is in excellent agreement with experimental data. 

While the SINDy framework has identified a single image vortex as the most parsimonious model, we emphasize that this does not negate the work of Ref.~\cite{Groszek2018}, whose suggestion of an infinite series of image vortices for a harmonic trap is physically well meaning. It does, however, provide a simple approximation for those working on vortex dynamics in different trapping geometries. This is an exciting step forward, as the ability to learn these equations from data not only allows for point vortex model simulations in power-law traps, it also removes the need to find a conformal mapping for given trap to a known image vortex case. In novel geometries emerging in ultra-cold atomic BECs, these mappings may be analytically intractable, so this is a powerful tool. Future applications of this method could include ring or double-ring geometries \cite{Bland_ring}, which have applications to the emerging field of atomtronics. Alternatively, adding dissipation to the model could lead to breakthroughs in understanding vortex-sound interactions in superfluids, a key question in superfluid turbulence where sound waves are expected to act as an energy sink in a quantum fluid as viscosity would in a classical fluid.

\newpage
\section*{Acknowledgments}
RD is grateful for stimulating discussions with Urban Fasel, Nick Keepfer, Matthew Juniper and Lloyd Fung at the 2023 and 2024 NFFDy Summer Programmes (held at Cambridge University and the University of Leeds, respectively), and to Andrew Baggaley and Nick Parker for discussion during the preparation of the manuscript. RD is supported by the UK Engineering and Physical Sciences Research Council, Grant No. EP/X028518/1. This work made use of the Rocket HPC Facility at Newcastle University.

\appendix
\section{The Gross-Pitaevskii Equation}
\label{appendix:GPE}

An ultra-cold gas of weakly interacting Bosons at zero-temperature is well described by a single macroscopic wavefunction, $\Psi(\rr,t)$. The dynamics of the wavefunction are determined by the Gross-Pitaevskii Equation (GPE),
\begin{equation}
    i\hbar\frac{\partial \Psi}{\partial t} = \left[ - \frac{\hbar^2}{2m}\nabla^2 + V(\rr) + g|\Psi|^2 - \mu \right] \Psi,
    \label{eqn:3d_gpe}
\end{equation}
where $m$ is the mass of the atomic species, $V$ is the trapping potential, $g$ parameterises the intra-species interactions of the gas, and $\mu$ is the chemical potential \cite{Primer}. Due to the high degree of experimental controllability, the gas can be made effectively 2D by applying a tight trapping potential in the $z$ axis, so
\begin{equation}
    V(\rr) = V_\perp(x,y) + \frac{1}{2}\omega_z^2 m z^2,
    \label{eqn:3d_trap}
\end{equation}
where the $\perp$ subscript denotes that the dependency is in the $xy$-plane. By tight, we require that $\hbar\omega_z \gg \mu$, which prevents excitations in the $z$ axis \cite{Rooney2011}. Under this assumption, we can write the 3D wavefunction as a product of the groundstate wavefunction of a harmonic trap in the $z$ axis and a 2D time-dependent wavefunction
\begin{equation}
    \Psi(\rr,t) = \frac{1}{\sqrt{\ell_z\sqrt{\pi}}} \exp\left( - \frac{z^2}{2\ell_z^2}\right) \psi_\perp (x,y,t).
    \label{eqn:3d_groundstate}
\end{equation}
Substituting Eqns.~\eqref{eqn:3d_trap} and \eqref{eqn:3d_groundstate} into Eqn.~\eqref{eqn:3d_gpe} and integrating over $z$ leads to the 2D GPE
\begin{equation}
    i\hbar\frac{\partial\psi_\perp}{\partial t} = \left[-\frac{\hbar^2}{2m}\nabla_\perp^2 + V_\perp(x,y) + g_{2D}|\psi_\perp|^2-\mu_{2D}\right] \psi_\perp,
\end{equation}
where $\ell_z^2=\hbar/m\omega_z$ is the harmonic oscillator length in $z$ and the 2D Laplacian term is $\nabla_\perp^2=\partial_x^2 + \partial_y^2$.  Here we have introduced two effective-2D parameters, the interaction parameter is 
\begin{equation}
    g_{2D} = \frac{2\sqrt{2\pi} \hbar^2 a_s}{m\ell_z},
\end{equation}
where $a_s$ is the $s$-wave scattering length of the atoms, and the 2D chemical potential is $\mu_{2D} = \mu - \frac{1}{2}\hbar\omega_z$.

Having introduced the 2D GPE, we proceed to write it in non-dimensional form. From this point, we will take the trapping potential to be of the form
\begin{equation}
    V_\perp(x,y) = \mu_{2D} \left(\frac{r}{R}\right)^{2p},
\end{equation}
where $r^2=x^2+y^2$ and $p$ is positive. The chemical potential is a characteristic energy scale of the system, and we can then introduce a characteristic length, $\ell_\perp = \hbar^2/m\mu_{2D}$, and a characteristic time, $\tau=\hbar/\mu_{2D}$. This allows us to re-cast the 2D GPE as 
\begin{equation*}
    i \frac{\partial \psi_\perp^\prime}{\partial t'} = \left[ -\frac{1}{2}\nabla^{\prime 2} + \left(\frac{r^\prime}{R^\prime}\right)^{2p} + g_{2D}^\prime |\psi_\perp^\prime|^2 -1 \right] \psi_\perp^\prime,
\end{equation*}
where primes denote non-dimensional quantities. This is Eqn.~\eqref{eqn:2D_GPE} in the main text. In the main text, and for the remainder of the appendix, we will drop the prime notation and the subscript $\perp$ where the meaning is clear. 

For a trapping potential of the form $V = \left(r/R\right)^{2p}$, we can approximate the density profile of the condensate as the Thomas-Fermi profile \cite{Primer} by neglecting the kinetic term in the 2D GPE. This leads to the profile
\begin{equation}
    \rho_{TF} = |\psi|^2 = \begin{cases} \dfrac{R^{2p} - r^{2p}}{g_{2D}R^{2p}} \qquad \text{when } r^2 \leq R^2, \\ \ \\  0 \qquad \qquad \ \ \ \text{ otherwise,} \end{cases} \label{eqn:TF_approximation}
\end{equation}
which has been experimentally shown to be a good approximation of the full condensate density solution \cite{Dalfovo1999}. We can then calculate the (non-dimensionalised) atom number, $N$ to be
\begin{equation}
    N = \int_0^{\vartheta=2\pi} \int_0^{r=R} \rho_{TF} \ r dr \, d\vartheta = \frac{\pi R^2}{g_{2D}} \frac{p}{p+1}.
\end{equation}

\section{The Image Vortex}
\label{appendix:image_vortex}
As discussed in Sec.~\ref{subsec:synthetic}, a vortex with charge $q_v$ and coordinates $\rr_v=\left(x_v,y_v\right)$ that is in a fluid contained in a infinitely hard-walled trap with radius $R$ will have an image vortex at $\rr_I = \rr_v R^2 / |\rr_v|^2$. Suppose, more generally, the image vortex is located at $\rr_I = \rr_v \varphi^2 / |\rr_v|^2$, where $\varphi$ is a parameter to be found. In the absence of dissipation, we can write the evolution of this vortex, following Eqn.~\eqref{eqn:general_pvm}, as
\begin{equation}
    \frac{d}{dt} \left( \begin{matrix} x_v \\ y_v \end{matrix}\right) = \frac{-q_v}{\left(x_v-x_I\right)^2+\left(y_v-y_I\right)^2}\left( \begin{matrix} - \left(y_v - y_I\right) \\ x_v - x_I \end{matrix} \right) - q_v \boldsymbol{\hat{z}}\times \nabla \ln \rho, 
\end{equation}
where $\rho$ is the background density of the fluid. We can calculate the distance between the vortex and the image vortex to be
\begin{equation}
    \left( \begin{matrix} x_v - x_I \\ y_v - y_i \end{matrix} \right) = \frac{x_v^2 + y_v^2 - \varphi^2}{x_v^2 + y_v^2}\left( \begin{matrix} x_v \\ y_v \end{matrix}\right),  
\end{equation}
yielding 
\begin{equation}
    \frac{d}{dt} \left( \begin{matrix} x_v \\ y_v \end{matrix}\right) = \frac{-q_v}{x_v^2 + y_v^2 - \varphi^2}\left( \begin{matrix} - y_v \\ x_v \end{matrix} \right) - q_v \boldsymbol{\hat{z}}\times \nabla \ln \rho. 
\end{equation}
We can approximate the background fluid away from the vortex core using the Thomas-Fermi approximation, Eqn.~\eqref{eqn:TF_approximation}, leading us to write
\begin{equation*}
    \nabla \ln \rho_\mathrm{TF}  = \frac{-2p \left(x^2 + y^2\right)^{p-1}}{R^{2p}-\left(x^2 + y^2\right)^p} \left(\begin{matrix} x \\ y \end{matrix}\right),
    \label{eqn:background_terms}
\end{equation*}
which is Eqn.~\ref{eqn:background_terms} in the main text. Hence, for a single vortex within the given trapping potential, the point vortex model reads
\begin{equation*}
    \frac{d}{dt} \left( \begin{matrix} x_v \\ y_v \end{matrix}\right) = \frac{-q_v}{x_v^2 + y_v^2 - \varphi^2}\left( \begin{matrix} - y_v \\ x_v \end{matrix} \right) - q_v \left(\begin{matrix} - w_y\left(x_v,y_v\right) \\ \quad w_x\left(x_v,y_v\right) \end{matrix} \right),
\end{equation*}
which is Eqn.~\eqref{eqn:power_trapped_vortex} in the main text. The Thomas-Fermi approximation depends only on the confining potential, which is static throughout the vortex evolution, and not on the position of the vortex. This allows us to re-write our equations of motion as 
    \begin{equation*}
    \begin{matrix}
        q_v y_v + \left( \dot{x}_v - q_v w_y\right) \left(\varphi^2 - x_v^2 - y_v^2\right) & = 0, \\
        q_v x_v - \left(\dot{y}_v + q_v w_x \right) \left(\varphi^2 - x_v^2 - y_v^2\right) & = 0,
    \end{matrix}
\end{equation*}
which is Eqn.~\eqref{eqn:point_vortex_fit} in the main text.



\bibliography{cas-refs}

\end{document}